\documentclass[onecolumn,showpacs,preprintnumbers,amsmath,amssymb,times,pre]{revtex4}

\usepackage{graphicx}% Include figure files
\usepackage{dcolumn}% Align table columns on decimal point
\usepackage{bm}% bold math
\usepackage{amsmath}
\usepackage{color}

\begin{document}

\title{Suppressing epidemic spreading by risk-averse migration in dynamical networks}

\author{Han-Xin Yang$^{1}$}\email{yanghanxin001@163.com}
\author{Ming Tang$^{2,3}$}
\author{Zhen Wang$^{4}$}

\affiliation{$^{1}$Department of Physics, Fuzhou University, Fuzhou
350116, China\\$^{2}$School of Information Science Technology, East
China Normal University, Shanghai 200062, China\\$^{3}$Big data
research center, University of Electronic Science and Technology of
China, Chengdu 611731, China\\$^{4}$Center for OPTical IMagery
Analysis and Learning (OPTIMAL), Northwestern Polytechnical
University, Xi'an 710072, Shaanxi, China}

\begin{abstract}

In this paper, we study the interplay between individual behaviors
and epidemic spreading in a dynamical network. We distribute agents
on a square-shaped region with periodic boundary conditions. Every
agent is regarded as a node of the network and a wireless link is
established between two agents if their geographical distance is
less than a certain radius. At each time, every agent assesses the
epidemic situation and make decisions on whether it should stay in
or leave its current place. An agent will leave its current place
with a speed if the number of infected neighbors reaches or exceeds
a critical value $E$. Owing to the movement of agents, the network's
structure is dynamical. Interestingly, we find that there exists an
optimal value of $E$ leading to the maximum epidemic threshold. This
means that epidemic spreading can be effectively controlled by
risk-averse migration. Besides, we find that the epidemic threshold
increases as the recovering rate increases, decreases as the contact
radius increases, and is maximized by an optimal moving speed. Our
findings offer a deeper understanding of epidemic spreading in
dynamical networks.
\end{abstract}

\date{\today}
\maketitle


\section{Introduction}

The study of epidemic spreading in complex networks has received
increasing attention since the 21st Century~\cite{rev}. In the
original model, individuals in the network are set to be ``passive''
nodes awaiting to be infected. However, in reality, when individuals
become aware of the potential disease, they would take preventive
measures (e.g., avoiding contact with infected individuals, wearing
protective masks, or vaccination) to protect themselves.

The interplay between individual behaviors and epidemic dynamics in
complex networks has attracted growing interest in recent
years~\cite{1,2,3,4,5,6,8,9,10,11,12}. For example, Gross $et$ $al.$
proposed an adaptive epidemic dynamical model, in which the
susceptible breaks the link to the infected and forms a new link to
another randomly selected susceptible, and the adaptive system
displayed assortative degree correlation, oscillations, hysteresis,
and first order transitions~\cite{Gross}. Funk $et$ $al.$ studied
the impacts of awareness spread on both epidemic threshold and
prevalence, and they found that, in a well-mixed population, spread
of awareness can reduce the prevalence of epidemic but does not tend
to affect the epidemic threshold~\cite{Funk}. Granell $et$ $al.$
investigated the interplay between awareness and epidemic spreading
in multiplex networks, and revealed the existence of a metacritical
point at which the critical onset of the epidemics starts depending
on the reaching of the awareness process~\cite{Granell}. Wang $et$
$al.$ studied the asymmetrically interacting spreading dynamics
based on a two susceptible-infected-recovered processes coupled
model in multiplex networks, and found that the propagation of
information can lead to the increase of the epidemic
threshold~\cite{Wang}. Xia $et$ $al.$ introduced a
Susceptible-Infected-Removed model with infection delay and
propagation vector~\cite{Xia1}. They found that both of the factors
can markedly reduce the critical threshold of disease infection and
accelerate the dynamical processes of epidemic spreading.
Furthermore, they studied effects of delayed recovery and nonuniform
transmission on the spreading of diseases~\cite{Xia2}.

It has been known that in real-life situations, when epidemics
outbreaks, individuals often escape from the infected areas and
migrate to other regions. Based on such fact, we propose a
risk-averse migration model, in which each individual assesses the
epidemic situation and make decisions on whether it should stay in
or leave its current place. We assume that an individual will
migrate to other places if the number of infected neighbors reaches
or exceeds a critical value of $E$ (we call it as the risk
threshold). The larger value of $E$ means that individuals are able
to endure a greater risk of being infected. Interestingly, we find a
nonmonotonic behavior in that the epidemic threshold can be
maximized at a moderate value of $E$. This means that we can control
the outbreak of epidemic by a proper choice of the risk-averse
migration.

\section{Model}

In our model, $N$ agents are distributed on a square-shaped cell of
size $L$ with periodic boundary conditions. We utilize the standard
susceptible-infected-susceptible (SIS) dynamical process for
modeling epidemic spreading. Two agents can contact with each other
if their distance is less than $r$ (which we call it as the contact
radius). An initial fraction of agents $\rho_{0}$ is set to be
infected (we set $\rho_{0}=0.1$ in numerical experiments). At each
time step, each susceptible agent is infected with probability
$\lambda$ if it contacts with an infected agent. At the same time,
infected agents are cured and become again susceptible with
probability $\mu$.

The initial positions of agents are randomly distributed. After each
time step, each agent $i$ decides whether to stay in or leave its
current position. If the number of infected agents within the
contact radius is less than a threshold $E$, an agent does not move.
Otherwise, agent $i$ moves and updates its position according to:
\begin{equation}
x_{i}(t+1)=x_{i}(t)+v\cos\theta_{i}(t),
 \label{1}
\end{equation}
\begin{equation}
y_{i}(t+1)=y_{i}(t)+v\sin\theta_{i}(t),
 \label{2}
\end{equation}
where $x_{i}(t)$ and $y_{i}(t)$ are the coordinates of agent $i$ at
time $t$, $v$ is the moving speed, and $\theta_{i}(t)$ is an random
variable chosen from the interval $[-\pi,\pi]$.

In the present model, we consider the local information-based
awareness, that is, an individual only knows the states of its
neighbors. Due to the limitation of local information, a migrant may
not wisely choose the new destination. Thus, the moving directions
are randomly set. Note that for $E=0$, all agents move at each time
step. For a sufficiently large value of $E$, all agents stay still.

\section{Results and discussions}

\begin{figure*}
\begin{center}
 \scalebox{1}[1]{\includegraphics{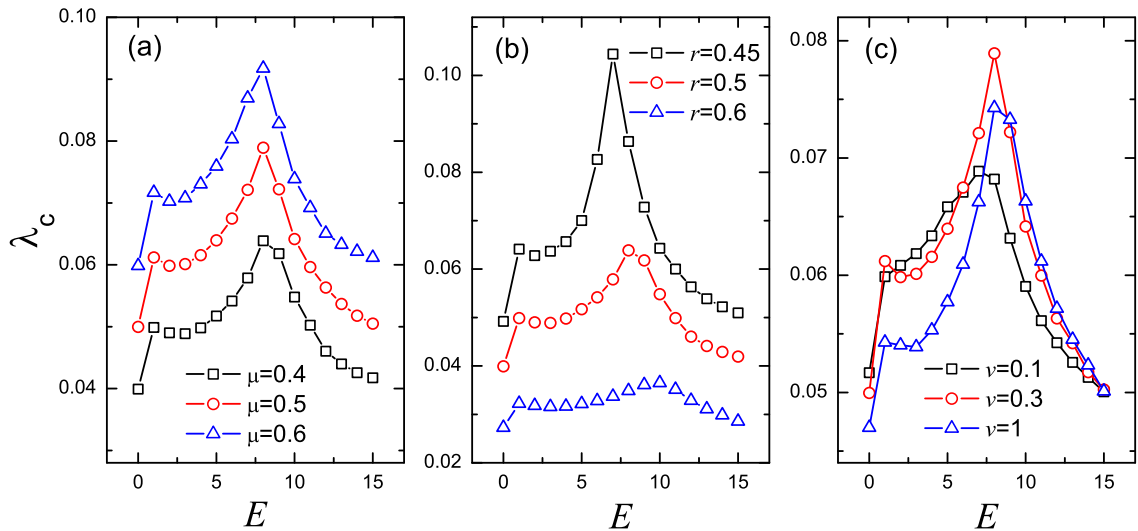}} \caption{(Color online)
(a) The epidemic threshold $\lambda_{c}$ as a function of the risk
threshold $E$ for different values of the recovering rate $\mu$. The
contact radius $r=0.5$ and the moving speed $v=0.3$. (b) The
epidemic threshold $\lambda_{c}$ as a function of $E$ for different
values of the contact radius $r$. The recovering rate $\mu=0.4$ and
the moving speed $v=0.3$. (c) The epidemic threshold $\lambda_{c}$
as a function of $E$ for different values of the moving speed $v$.
The recovering rate $\mu=0.5$ and the contact radius $r=0.5$. Each
data point is result of averaging over 100 random realizations.}
\label{fig1}
\end{center}
\end{figure*}

In all the following simulations, we set the total number of agents
$N = 1500$ and the size of the square region $L = 10$ if there is no
special mention.

A fundamental quantity in SIS dynamics is the epidemic threshold
$\lambda_{c}$, below which the epidemic dies out. Due to the
complexity of the model, it is very hard for us to obtain an
analytical expression of $\lambda_{c}$. However, for large values of
$v$ and $E=0$, the population is well mixed and the epidemic
threshold can be given as:
\begin{equation}
\lambda_{c}=\frac{\mu}{\langle k \rangle} = \frac{\mu L^{2}}{\pi
r^{2}N},
 \label{3}
\end{equation}
where $\langle k \rangle$ denotes the average number of neighbors at
each time step.

Figure~\ref{fig1} shows the dependence of $\lambda_{c}$ on the risk
threshold $E$ for different values of the recovery rate $\mu$, the
contact radius $r$ and the moving speed $v$. From Fig.~\ref{fig1},
one can see that for given values of other parameters, there exists
an optimal value of $E$, hereafter denoted by $E_{opt}$, resulting
in the maximum $\lambda_{c}$.

\begin{figure}
\begin{center}
 \scalebox{0.75}[0.75]{\includegraphics{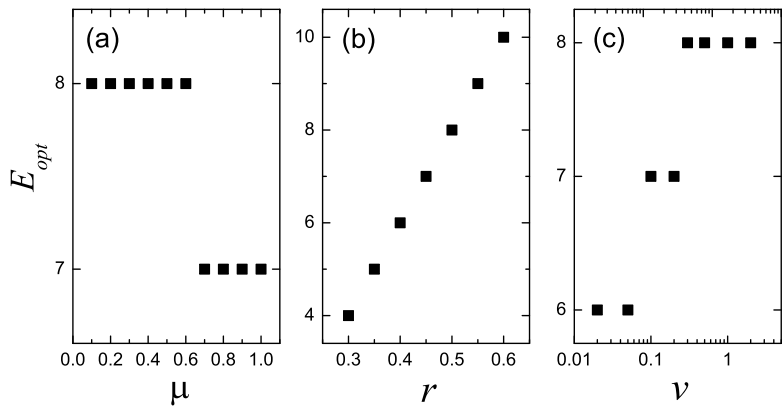}}
\caption{(a) The optimal value of the risk threshold $E$, $E_{opt}$,
as a function of the recovering rate $\mu$. The contact radius
$r=0.5$ and the moving speed $v=0.3$. (b) The dependence of
$E_{opt}$ on the contact radius $r$. The recovering rate $\mu=0.4$
and the moving speed $v=0.3$. (c) The dependence of the risk
threshold $E_{opt}$ on the moving speed $v$. The recovering rate
$\mu=0.5$ and the contact radius $r=0.5$. } \label{fig2}
\end{center}
\end{figure}

\begin{figure}
\begin{center}
 \scalebox{0.75}[0.75]{\includegraphics{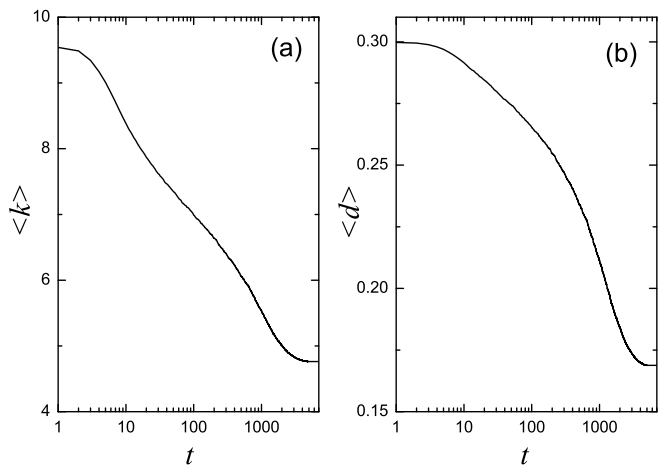}}
 \caption{(a) The average number of neighbors $\langle k \rangle$ and (b) the average distance of
two neighboring agents $\langle d \rangle$, as time $t$ evolves. The
spreading rate $\lambda=0.15$, the recovering rate $\mu=0.4$, the
contact radius $r=0.45$, the moving speed $v=0.3$, and the risk
threshold $E=7$. Each curve is result of averaging over 100 random
realizations.}
 \label{fig3}
\end{center}
\end{figure}

Figure~\ref{fig2} shows the risk threshold $E_{opt}$ as a function
of the recovering rate $\mu$, the contact radius $r$ and the moving
speed $v$ respectively. For given values of other parameters,
$E_{opt}$ decreases from 8 to 7 as $\mu$ increases from 0.1 to 1
[see Fig.~\ref{fig2}(a)], $E_{opt}$ increases from 4 to 10 as $r$
increases from 0.3 to 0.6 [see Fig.~\ref{fig2}(b)], and $E_{opt}$
increases from 6 to 8 as $v$ increases from 0.02 to 2 [see
Fig.~\ref{fig2}(c)].

For very large values of $E$, agents do no move and they have no
chances to escape from the attack of infected neighbors. For very
small values of $E$ (e.g., $E=0$), almost all agents move with the
random directions, which cannot reduce the probability that
encounters an infected agent. For the moderate value of $E$, agents
will leave the heavily infected areas and stay at the slightly
infected regions. An agent with more neighbors has higher
probability of being infected and it tends to leave. On the other
hand, an agent with less neighbors is more likely to stay at its
current place as the number of infected neighbors is less than $E$.
As a result, the average number of neighbors $\langle k \rangle$
gradually decreases due to the migration driven by the moderate
value of $E$ [see Fig.~\ref{fig3}(a)].

Counterintuitively, the decrease of the number of neighbors does not
enlarge the distance between the two neighboring agents. From
Fig.~\ref{fig3}(b), we are surprising to find that the average
distance of two neighboring agents $\langle d \rangle$ decreases as
time $t$ evolves for the moderate value of $E$. This abnormal
phenomenon indicates that agents are spontaneously divided into
different components under the moderately risk-averse migration.

From the perspective of complex network theory, the topology of the
system can be expressed as a dynamical network of mobile
agents~\cite{dynamical4,dynamical5,dynamical6,dynamical7,dynamical8}.
Every agent is regarded as a node of the network and a wireless link
is established between two agents if their geographical distance is
less than the contact radius $r$. Owing to the movement of agents,
the network's structure changes with time.  The decrease of the
average number of neighbors and the average distance of two
neighboring nodes indicates that agents are divided into different
components under the moderately risk-averse migration. A component
is a subgraph in which any two nodes are connected by a path. Note
that each component is temporal as agents move with time.

\begin{figure}
\begin{center}
 \scalebox{0.75}[0.75]{\includegraphics{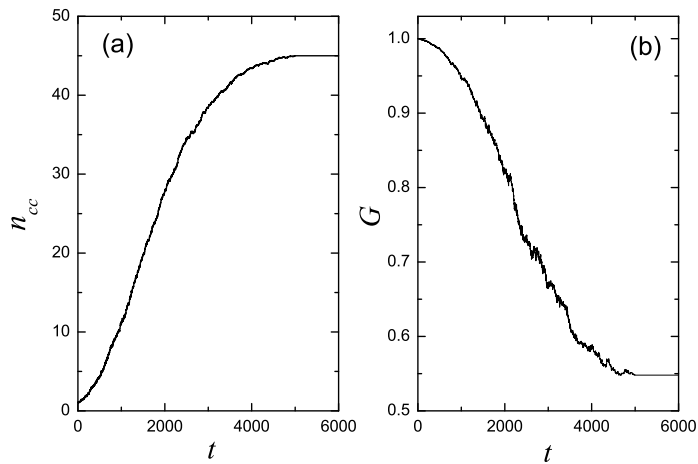}}
 \caption{(a) The number of connected components $n_{cc}$ and (b) the normalized size of
the largest component $G$, as time $t$ evolves.  The spreading rate
$\lambda=0.15$, the recovering rate $\mu=0.4$, the contact radius
$r=0.45$, the moving speed $v=0.3$, and the risk threshold $E=7$.
Each curve is result of averaging over 100 random realizations.}
\label{fig4}
\end{center}
\end{figure}

Figure~\ref{fig4} shows the number of connected components $n_{cc}$
and the normalized size of the largest component $G$ as time $t$
evolves when the risk threshold $E$ is moderate. Initially, all
agents form a connected component, corresponding to $n_{cc}=1$ and
$G=1$. As time evolves, the system is divided into more and more
components [see Fig.~\ref{fig4}(a)] and the size of the largest
component decreases [see Fig.~\ref{fig4}(b)].

Combining Figs.~\ref{fig3} and ~\ref{fig4}, we can understand why
the epidemic spreading can be most inhibited by the moderately
risk-averse migration. For non-migration ($E$ is very large) and
constant moving ($E$ is very small), all agents construct a
connected component and the average number of neighbors keep almost
unchanged over time (the results are not shown here). For the
moderate value of $E$, the formation of different components stop
the epidemic spreading from one region to other regions. Moreover,
for the moderately risk-averse migration, the average number of
neighbors becomes smaller as time evolves, which greatly reduces the
risk of being infected. Both of the above factors lead to the
maximum epidemic threshold for the moderate value of $E$.

\begin{figure*}
\begin{center}
 \scalebox{0.85}[0.85]{\includegraphics{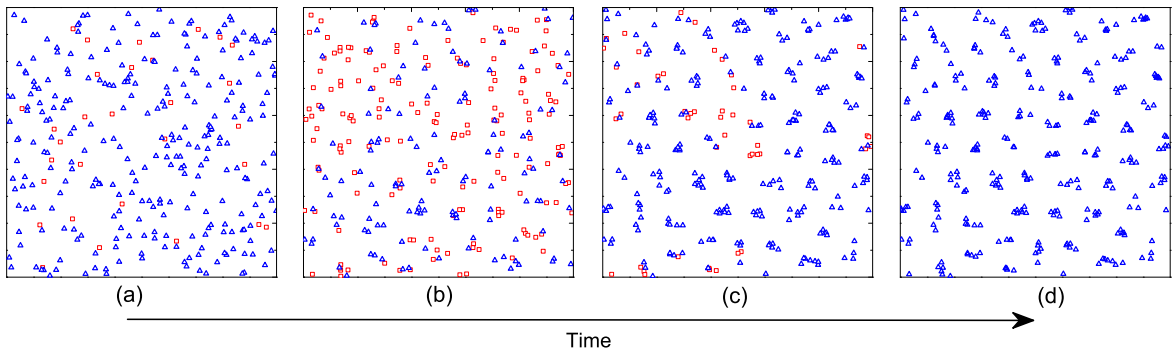}} \caption{(Color online)
Snapshots of spatial distributions of susceptible agents (denoted by
blue triangle) and infected agents (denoted by red square) at
different time steps. The number of agents $N=300$, the size of the
square region $L = 10$, the spreading rate $\lambda= 0.17$, the
recovering rate $\mu= 0.4$, the contact radius $r = 1$, the moving
speed $v = 0.3$, and the risk threshold $E = 7$.} \label{fig5}
\end{center}
\end{figure*}

\begin{figure*}
\begin{center}
 \scalebox{0.75}[0.75]{\includegraphics{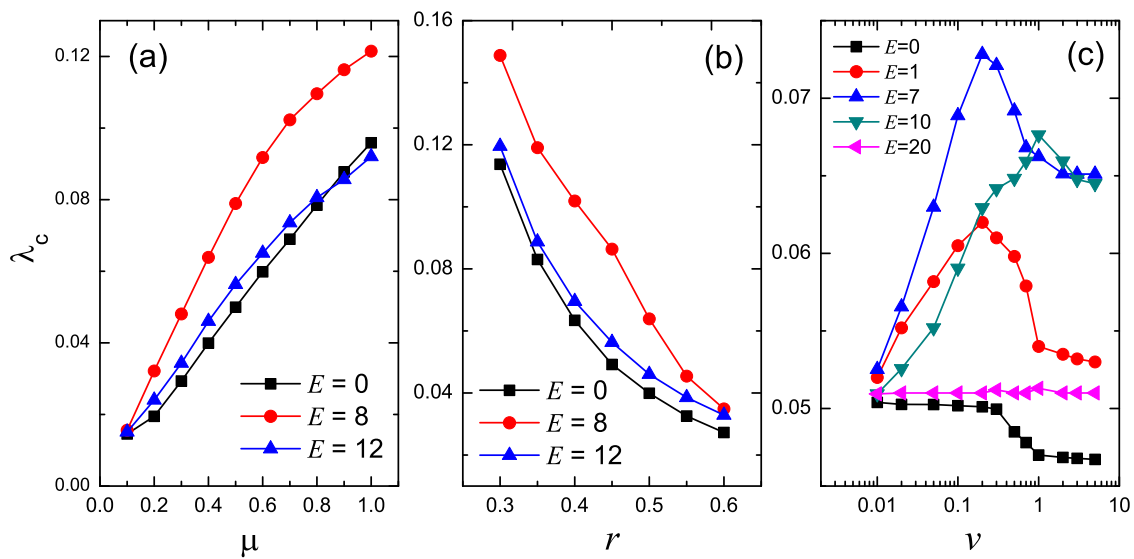}} \caption{(Color online) (a) The epidemic threshold $\lambda_{c}$ as a function of the recovering rate $\mu$
for different values of the risk threshold $E$. The contact radius
$r=0.5$ and the moving speed $v=0.3$. (b) The epidemic threshold
$\lambda_{c}$ as a function of the contact radius $r$ for different
values of $E$. The recovering rate $\mu=0.4$ and the moving speed
$v=0.3$. (c) The epidemic threshold $\lambda_{c}$ as a function of
the moving speed $v$ for different values of $E$. The recovering
rate $\mu=0.5$ and the contact radius $r=0.5$. Each data point is
result of averaging over 100 random realizations.} \label{fig6}
\end{center}
\end{figure*}

To intuitively understand how the moderate value of $E$ affects the
process of epidemic spreading, we plot the spatial distribution of
susceptible and infected agents evolves with time when the spreading
rate is a little smaller than the epidemic threshold. Initially, a
small fraction of infected agents is randomly distributed [see Fig.
5(a)]. After a few time steps, the density of infected agents
increases [see Fig. 5(b)]. However, with time the number of
susceptible agents continually increases [see Fig. 5(c)] and finally
they take over the whole system [see Fig. 5(d)]. Combining Figs.
5(a) and 5(d), one can observe that agents eventually aggregate to
many different components in which they stay more closely than that
in the initial state.

Finally, we investigate the effects of the recovery rate $\mu$, the
contact radius $r$ and the moving speed $v$ on epidemic spreading.
From Fig.~\ref{fig6}(a), we see that for a given value of $E$, the
epidemic threshold $\lambda_{c}$ increases with $\mu$, which is in
accordance with previous studies~\cite{rev}. The epidemic threshold
$\lambda_{c}$ decreases with $r$ [see Fig.~\ref{fig6}(b)]. In our
model, the smaller value of $r$ corresponds to the decrease of the
neighbors. As is well known, the reduction in the number of the
contacting neighbors effectively suppresses epidemic spreading,
leading to a larger epidemic threshold~\cite{ave}. From
Fig.~\ref{fig6}(c), one can see that the epidemic threshold
$\lambda_{c}$ decreases with $v$ when $E=0$. For very large values
of $E$ (e.g., $E=20$), almost all agents do not move. In this case,
the epidemic threshold does not change as the moving speed
increases. For moderate values of $E$, the relation between
$\lambda_{c}$ and $v$ is nonmonotonic, where an optimal value of $v$
can maximize the epidemic threshold. From Fig.~\ref{fig6}(c), one
can also observe that the optimal value of $v$ increases as $E$
increases. For small values of $v$, agents move so slowly that the
system is very close to the static network in which each agent's
neighbors keep unchanged. For large values of $v$, agents move so
fast that they cannot stay close to each other and cannot be divided
into different components. Therefore, there must be an intermediate
value of $v$ that maximizes the epidemic threshold.

\section{Conclusions}

To summarize, we have studied the effects of risk-averse migration
on epidemic spreading. Each agent has the same contact radius $r$.
Two agents can contact with each other if the geographical distance
between them is less than $r$. An agent will move to a randomly
chosen place with the speed $v$ if the number of infected agents
within its contact radius reaches or exceeds a risk threshold $E$.
We find that the epidemic threshold can be maximized by an optimal
value of $E$. For the moderate value of $E$, agents gradually form
different components in which the average number of neighbors is
much smaller than that for very small or very large value of $E$.

The interplay between information awareness and epidemic spreading
in dynamical networks composed by mobile agents is an interesting
research topic. Our findings presented here raise a number of open
questions, answers to which could further deepen our understanding
of the role of migration in epidemic spreading. For example, what
happens if the contact radius or the moving speed is not the same
for all agents? Another interesting issue is whether agents can move
with the preferential directions rather than the random directions?
In the present work, an individual determinately leaves if the
number of infected neighbors reaches or exceeds E. What would happen
if an individual leaves with some probability?

\begin{acknowledgments}
This work was supported by the National Science Foundation of China
(Grants No.~61403083, No.~71301028 and No.~11575041).
\end{acknowledgments}

\end{document}